\documentclass[letterpaper, paper,11pt]{AAS}		

\usepackage{AAS_packages}

\PaperNumber{17--720}

\begin{document}

\title{Geometric Control for Autonomous Landing on Asteroid Itokawa using visual localization}

\author{Shankar Kulumani, Kuya Takami, and Taeyoung Lee\thanks{Mechanical and Aerospace Engineering, George Washington University, 800 22nd St NW, Washington, DC 20052, Tel: 202--994--8710, Email: \href{mailto:skulumani@gwu.edu}{\{skulumani,kuya,tylee\}@gwu.edu}.}
}

\maketitle{} 		

\begin{abstract}
    This paper considers the coupled orbit and attitude dynamics of a dumbbell spacecraft around an asteroid. 
    Geometric methods are used to derive the coupled equations of motion, which are defined on the configuration space of the special Euclidean group, and then a nonlinear controller is designed to enable trajectory tracking of desired landing trajectories.
    Rather than relying on sliding mode control or optimization based methods, the proposed approach avoids the increased control utilization and computational complexity inherent in other techniques.
    The nonlinear controller is used to track a desired landing trajectory to the asteroid surface. 
    A monocular imaging sensor is used to provide position and attitude estimates using visual odometry to enable relative state estimates.
    We demonstrate this control scheme with a landing simulation about asteroid Itokawa.
\end{abstract}

\section{Introduction}\label{sec:introduction}
Small solar system bodies, such as asteroids and comets, are of significant interest to the scientific community.
These small bodies offer great insight into the early formation of the solar system.
This insight offers additional detail into the formation of the Earth and also the probable formation of other extrasolar planetary bodies.
Of particular interest are those near-Earth asteroids (NEA) which inhabit heliocentric orbits in the vicinity of the Earth.
These easily accessible bodies provide attractive targets to support space industrialization, mining operations, and scientific missions.
NEAs potentially contain many materials such as those useful for propulsion, construction, or for use in semiconductors.
Also, many bodies contain highly profitable materials, such as precious or strategic metals that can support a new space focused market~\cite{ross2001}.

Furthermore, these asteroids are of keen interest for more practical purposes.
The recent meteor explosions in  2002 over Tagish Lake, Canada or over Chelyabinsk, Russia in 2013 are clear evidence of the risk of asteroid impacts on the Earth.
These asteroids, which released an energy equivalent to \SI{5}{\kilo\tonne} of TNT, are estimated to strike the Earth on average every year~\cite{brown2002}.
Larger bodies, such as the \SI{60}{\meter} object that exploded over Tunguska, Russia in 1908, release the energy equivalent to \SI{10}{\mega\tonne} of TNT and will occur on average every \num{1000} years.
Asteroids and comets are the greatest threat to future civilizations and as a result there is a focused effort to mitigate these risks~\cite{wie2008}.
A wide variety of strategies, including nuclear standoff detonation, mass drivers, kinetic-energy projectiles, and low-thrust deflection via electric propulsion or solar sails, have been proposed to deal with the technically challenging asteroid mitigation problem~\cite{adams2004}.
In spite of the significant interest in asteroid deflection, and the extensive research by the community, the operation of spacecraft in their vicinity remains a challenging problem.

While there has been significant study of interplanetary transfer trajectories, relatively less analysis has been conducted on operations in the vicinity of asteroids.
The dynamic environment around asteroids is strongly perturbed and challenging for analysis and mission operations~\cite{scheeres1994,scheeres2000}.
Due to their low mass, which results in a low gravitational attraction, asteroids may have irregular shapes and potentially chaotic spin states.
As a result, typical approaches of assuming a inverse square gravitational model are at best inaccurate and at worst do not capture the true dynamic environment.
In addition, the vast majority of asteroid are difficult to track or measure using current ground-based optical sensors. 
Due to their small size, frequently less than \SI{1}{\kilo\meter}, and low albedo the reflected energy of these asteroids is insufficient for reliable detection or tracking.
Therefore, the dynamic model of the asteroid is relatively coarse prior to arrival of a dedicated spacecraft in the vicinity. 
As a result, any spacecraft mission to an asteroid is dependent on a robust dynamic simulation and must incorporate the ability to deal with uncertain forces and environments.

Furthermore, since the magnitude of the gravitational attraction is relatively small, non-gravitational effects, such as solar radiation pressure or third-body effects, become much more significant.
As a result, the orbital environment is generally quite complex and it is difficult to generate analytical insights.
One key consideration is the coupling between rotational and translational states around the asteroid.
The coupling is induced due to the different gravitational forces experienced on various parts of the spacecraft.
The effect of the gravitational coupling is related to the parameter \(\epsilon = \frac{r}{R_c}\), where \(r\) is the characteristic spacecraft length and \(R_c\) is the orbital radius~\cite{hughes2004}.
For Earth based missions, the orbital radius is several orders of magnitude larger than the spacecraft length and \(\epsilon\) is small.
As a result, the corresponding gravitational moment is weak and can be neglected. 
Therefore, the translational and rotational equations of motion become decoupled and can be considered separately, significantly simplifying the analysis. 
However, for operations around an asteroid the orbital radius is much smaller, which leads to much larger values of \(\epsilon\) and much larger influence of the rotational and translational coupling.
References~\citenum{elmasri2005} and~\citenum{sanyal2004} investigated the coupling of an elastic dumbbell spacecraft in orbit about a central body, but only considered the case of a spherically symmetric central body.
Furthermore, the spacecraft model is assumed to remain in a planar orbit.
As result, these developments are not directly applicable to motion about an asteroid, which experience highly non-keplerian dynamics.

An additional layer of complexity is the design of landing trajectories on asteroids.
Beginning with the first landing of NEAR Shoemaker on asteroid 433 Eros, there has been a concerted effort to develop techniques and methodologies for asteroid landing~\cite{dunham2002, kubota2006}.
There is already considerable knowledge on the planetary landing problem~\cite{acikmese2007, meditch1964, ingoldby1978}.
While conceptually similar, the landing of spacecraft on small bodies requires additional consideration. 
The surface of an asteroid is highly irregular and, as discussed previously, there is a large coupling between the translational and rotational dynamics of the vehicle, which is further exaggerated when close to the surface.
References~\citenum{guelman1994, furfaro2013, zexu2012} consider the soft landing problem on an asteroid.
These approaches were primarily based on nonlinear control techniques which allowed for the development of closed loop controllers which enable landing.
However, only the translational dynamics of the body was considered and no notion of the attitude dynamics or it's coupling to the position is considered.
Furthermore, relatively simple gravitational models are used which make the results unsuitable for operations near irregular bodies.

In this paper, we develop a landing scheme for spacecraft on an asteroid.
The main objective is to construct the coupled equations of motion of a rigid spacecraft about an asteroid.
This accurate dynamic model is then used to derive a nonlinear controller for the tracking of a landing trajectory.
In contrast to much of the previous work, we explicitly consider the gravitational coupling between the orbit and attitude dynamics.
In addition, we utilize a polyhedron potential model to represent the shape of the asteroid, which results in an exact closed form expression of the gravitational potential field~\cite{werner1994,werner1996}.
This type of potential model is exact given the accuracy of the shape model and valid at all point outside of the body. 
As a result, the polyhedron model is ideal for all phases of spacecraft operations, from arrival to landing.

Determination of the state of the spacecraft is typically dependent on ground based observations with sporadic state updates.
An additional capability is possible by using on-board sensors to estimate the state of the spacecraft relative to the asteroid~\cite{kubota2003,miso1999}.
We seek to incorporate localization using monocular images into the geometric nonlinear controller. 
This will offer the ability to increase the accuracy and frequency of state estimates and allow for closed loop control for landing operations. 
The use of imagery also enables the use of computer vision algorithms and tools to estimate the shape and motion of the asteroid.
The combination of visual localization and geometric control is a new avenue for the operation of spacecraft near asteroids.

In short, this paper presents a nonlinear controller for the coupled motion of a spacecraft around an asteroid.
The dynamics are developed on the nonlinear manifold of rigid body motions, namely the special euclidean group.
This intrinsic geometric formulation accurately captures the coupling between the orbit and attitude dynamics. 
Due to the relative size of the spacecraft as compared to the orbital radius, there is a significant gravitational moment on the spacecraft. 
Through the use of the polyhedron gravitational model we ensure an accurate representation of the gravitational moment on the spacecraft throughout all phases of flight. 
Furthermore, we present a nonlinear controller developed on the special euclidean group which asymptotically tracks a desired landing trajectory. 

\section{Mathematical Formulation}\label{se:mathematical_problem}
In this paper, we consider the landing of a dumbbell model of a spacecraft onto an asteroid.
The dumbbell spacecraft consists of two masses connected by a massless rod and is a well-known representation of a multi body spacecraft.
Furthermore, the dumbbell model captures the important interactions of the coupling between orbital and attitude dynamics. 
As a result, this simple model is useful to capture the main characteristics of a wide variety of spacecraft configurations.
Typically, spacecraft have mass concentrated in a central structure, referred to as the bus, which houses the command and control system, actuators, fuel, sensors etc. 
In addition, comparatively light-weight solar panels extend from the bus to provide electrical energy from solar radiation. 
As a result, the distributed mass of the spacecraft is captured with the dumbbell representation.
In this section, we briefly review the polyhedron potential model and then present the derivation of the coupled dynamics of a dumbbell spacecraft about an asteroid.

\subsection{Polyhedron Potential Model}\label{sec:polyhedron_potential}

An accurate gravitational potential model is necessary for the operation of spacecraft about asteroids.
Additionally, a detailed shape model of the asteroid is needed for trajectories passing close to the body.
The classic approach is to expand the gravitational potential into a harmonic series and compute the series coefficients.
However, the harmonic expansion is always an approximation as a result of the infinite order series used in the representation.
Additionally, the harmonic model used outside of the circumscribing sphere is not guaranteed to converge inside the sphere, which makes it unsuitable for trajectories near the surface.

We represent the gravitational potential of the asteroid using a polyhedron gravitation model.
This model is composed of a polyhedron, which is a three-dimensional solid body, that is defined by a series of vectors in the body-fixed frame.
The vectors define vertices in the body-fixed frame as well as planar faces which compose the surface of the asteroid.
We assume that each face is a triangle composed of three vertices and three edges.
As a result, only two faces meet at each edge while three faces meet at each vertex.
Only the body-fixed vectors, and their associated topology, is required to define the exterior gravitational model.
References~\citenum{werner1994} and~\citenum{werner1996} give a detailed derivation of the polyhedron model.
Here, we summarize the key developments and equations required for implementation.

Consider three vectors \( \vecbf{v}_1, \vecbf{v}_2, \vecbf{v}_3 \in \R^{3 \times 1} \), assumed to be ordered in a counterclockwise direction about an outward facing normal vector, which define a face.
It is easy to define the three edges of each face as
\begin{align}\label{eq:edges}
    \vecbf{e}_{i+1,i} = \vecbf{v}_{i+1} - \vecbf{v}_i \in \R^{3 \times 1 },
\end{align}
where the index \( i \in \parenth{1,2,3} \) is used to permute all edges of each face.
Since each edge is a member of two faces, there exist two edges which are defined in opposite directions between the same vertices.
We can also define the outward normal vector to face \( f\)  as
\begin{align}\label{eq:face_normal}
    \hat{\vecbf{n}}_f &= \parenth{\vecbf{v}_{2} - \vecbf{v}_1} \times \parenth{\vecbf{v}_{3} - \vecbf{v}_2} \in \R^{3 \times 1},
\end{align}
and the outward facing normal vector to each edge as
\begin{align}\label{eq:edge_normal}
    \hat{\vecbf{n}}_{i+1,i}^f &= \parenth{\vecbf{v}_{i+1} - \vecbf{v}_i} \times \hat{\vecbf{n}}_f \in \R^{3 \times 1}.
\end{align}
For each face we define the face dyad \( \vecbf{F}_f \) as
\begin{align}\label{eq:face_dyad}
    \vecbf{F}_f &= \hat{\vecbf{n}}_f \hat{\vecbf{n}}_f \in \R^{3 \times 3}.
\end{align}
Each edge is a member of two faces and has an outward pointing edge normal vector, given in~\cref{eq:edge_normal}, perpendicular to both the edge and the face normal.
For the edge connecting the vectors \( \vecbf{v}_1 \) and \( \vecbf{v}_2 \), which are shared between the faces \(A\) and \( B\), the per edge dyad is given by
\begin{align}\label{eq:edge_dyad}
    \vecbf{E}_{12} = \hat{\vecbf{n}}_A \hat{\vecbf{n}}_{12}^A + \hat{\vecbf{n}}_B \hat{\vecbf{n}}_{21}^B \in \R^{3 \times 3}.
\end{align}
The edge dyad \( \vecbf{E}_e  \), is defined for each edge and is a function of the two adjacent faces meeting at that edge.
The face dyad \( \vecbf{F}_f \), is defined for each face and is a function of the face normal vectors.

Let \( \vecbf{r}_i \in \R^{3 \times 1} \) be the vector from the spacecraft to the vertex \( \vecbf{v}_i \) and it's length is given by \( r_i = \norm{\vecbf{r}_i} \in \R^{1} \).
The per-edge factor \( L_e \in \R^{1}\), for the edge connecting vertices \( \vecbf{v}_i \) and \( \vecbf{v}_j \), with a constant length \( e_{ij} = \norm{\vecbf{e}_{ij}} \in \R^1\) is
\begin{align}\label{eq:edge_factor}
    L_e &= \ln \frac{r_i + r_j + e_{ij}}{r_i + r_j - e_{ij}}.
\end{align}
For the face defined by the vertices \( \vecbf{v}_i, \vecbf{v}_j, \vecbf{v}_k \) the per-face factor \( \omega_f \in \R^{1} \) is
\begin{align}\label{eq:face_factor}
    \omega_f &= 2 \arctan \frac{\vecbf{r}_i \cdot \vecbf{r}_j \times \vecbf{r}_k}{r_i r_j r_k + r_i \parenth{\vecbf{r}_j \cdot \vecbf{r}_k} + r_j \parenth{\vecbf{r}_k \cdot \vecbf{r}_i} + r_k \parenth{\vecbf{r}_i \cdot \vecbf{r}_j}}.
\end{align}
The gravitational potential due to a constant density polyhedron is given as
\begin{align}\label{eq:potential}
    U(\vecbf{r}) &= \frac{1}{2} G \sigma \sum_{e \in \text{edges}} \vecbf{r}_e \cdot \vecbf{E}_e \cdot \vecbf{r}_e \cdot L_e - \frac{1}{2}G \sigma \sum_{f \in \text{faces}} \vecbf{r}_f \cdot \vecbf{F}_f \cdot \vecbf{r}_f \cdot \omega_f \in \R^1,
\end{align}
where \( \vecbf{r}_e\) and \(\vecbf{r}_f \) are the vectors from the spacecraft to any point on the respective edge or face, \( G\) is the universal gravitational constant, and \( \sigma \) is the constant density of the asteroid.
Furthermore we can use these definitions to define the attraction, gravity gradient matrix, and Laplacian as
\begin{align}
    \nabla U ( \vecbf{r} ) &= -G \sigma \sum_{e \in \text{edges}} \vecbf{E}_e \cdot \vecbf{r}_e \cdot L_e + G \sigma \sum_{f \in \text{faces}} \vecbf{F}_f \cdot \vecbf{r}_f \cdot \omega_f \in \R^{3 \times 1} , \label{eq:attraction}\\
    \nabla \nabla U ( \vecbf{r} ) &= G \sigma \sum_{e \in \text{edges}} \vecbf{E}_e  \cdot L_e - G \sigma \sum_{f \in \text{faces}} \vecbf{F}_f \cdot \omega_f \in \R^{3 \times 3}, \label{eq:gradient_matrix}\\
    \nabla^2 U &= -G \sigma \sum_{f \in \text{faces}}  \omega_f \in \R^1.\label{eq:laplacian}
\end{align}

One interesting thing to note is that both~\cref{eq:face_dyad,eq:edge_dyad} can be precomputed without knowledge of the position of the satellite.
They are both solely functions of the vertices and edges of the polyhedral shape model and are computed once and stored.
Once a position vector \( \vecbf{r} \) is defined, the scalars given in~\cref{eq:edge_factor,eq:face_factor} can be computed for each face and edge.
Finally,~\cref{eq:potential} is used to compute the gravitational potential on the spacecraft.
The Laplacian, defined in~\cref{eq:laplacian}, gives a simple method to determine if the spacecraft has collided with the body~\cite{werner1996}. 
\subsection{Dumbbell Spacecraft Equations of Motion}\label{sec:dumbbell}

The configuration space for rigid body motion is the semi-direct product, \(\SE = \R^3 \times \SO \), namely the special euclidean group.
The variations should be carefully constructed such that they respect the geometry of the configuration space.
By expressing the motion of the dumbbell directly on the special euclidean group, we avoid the issues inherent in using other kinematic representations which fail to preserve the geometric properties of the configuration space.
The kinematics of the dumbbell and asteroid are described in the inertial frame by
\begin{itemize}
    \item \( \vecbf{x} \in \R^3 \): the position of the center of mass of the dumbbell spacecraft represented in the inertial frame \( \vecbf{e}_i\)
    \item \( R \in \SO\): the rotation matrix which transforms vectors defined in the spacecraft fixed frame, \( \vecbf{b}_i \), to the inertial frame, \( \vecbf{e}_i \)
    \item \( \vecbf{\Omega} \in \R^3 \): the angular velocity of the spacecraft body fixed frame relative to the inertial frame and represented in the dumbbell body fixed frame \( \vecbf{b}_i \)
    \item \( R_A \in \SO \): the rotation matrix which transforms vectors defined in the asteroid fixed frame, \( \vecbf{f}_i \), to the inertial frame, \( \vecbf{e}_i \)
\end{itemize}
In this work, we assume that the asteroid is much more massive than the spacecraft and its motion is not affected by that of the spacecraft.
This assumption allows us to treat the motion of the vehicle independently from that of the asteroid, instead of treating the more complicated full-body problem. 

Using our kinematic variables we can define the kinetic and potential energy of the dumbbell as
\begin{align}\label{eq:kinetic_energy}
    T &= \frac{1}{2} m \norm{\dot{\vb{x}}}^2 + \frac{1}{2} \tr{S(\vb{\Omega}) J_d S\parenth{\vb{\Omega}}^T} , \\
    V( \vecbf{x}, R ) &=  - m_1 U \parenth{R_A^T \parenth{\vecbf{x} + R \vecbf{\rho}_1}} - m_2 U \parenth{R_A^T \parenth{\vecbf{x} + R \vecbf{\rho}_2}} ,
\end{align}
where the polyhedron potential is defined in~\cref{eq:potential}.
The position of each mass \(m_i\) of the dumbbell is defined in the dumbbell fixed frame by the vector \(\vb{\rho}_i\). 
The next step is to define the variations of the kinetic and potential energy to derive the equations of motion, which are given as
\begin{align} 
    \delta V &= -\sum_{i=1}^2  m_i \parenth{R_A \deriv{U}{\vb{z}_i} }^T \delta \vb{x} + m_i \hat{\vb{\eta}}\cdot \hat{\vb{\rho}_1} R^T R_A \deriv{U}{\vb{z}_i}, \\
    \delta T &= \parenth{m_1 + m_2} \dot{\vecbf{x}}^T \delta \dot{\vb{x}} + \frac{1}{2} \tr{- \dot{\hat{\vb{\eta}}} S(J \vb{\Omega}) + \hat{\vb{\eta}} S(\hat{\vb{\Omega}} J \vb{\Omega})}. 
\end{align}

Using the variations of the kinetic and potential energy we can derive the equations of motion of the dumbbell spacecraft about an asteroid using Hamilton's principle. 
Hamilton's principle then states that the variation of the action integral
\begin{align}
    \mathsf{G} = \int_{t_0}^{t_f} T(\dot{q}) - V(q) dt,
\end{align}
is stationary with fixed endpoints. 
Applying the calculus of variations and integration by parts results in the familiar Euler-Lagrange equations of motion.
Applying the Legendre transformation allows for the same dynamics to be expressed in an equivalent form as Hamilton's equations~\cite{lanczos1970}.
The equations of motion of a dumbbell spacecraft influenced by a polyhedron potential model are given as
\begin{align}
    \dot{\vb{x}} &= \vb{v}, \label{eq:position_kinematics}\\
    \parenth{m_1 + m_2} \dot{\vecbf{v}} &= m_1 R_A \deriv{U}{\vecbf{z}_1} + m_2 R_A \deriv{U}{\vecbf{z}_2} + \vecbf{u}_f, \label{eq:translational_dynamics}\\
    \dot{R} &= R S(\vb{\Omega}) , \label{eq:attitude_kinematics}\\
    J \dot{\vecbf{\Omega}} + \vecbf{\Omega} \times J \vecbf{\Omega} &= \vecbf{M}_1 + \vecbf{M}_2 + \vecbf{u}_m. \label{eq:attitude_dynamics}
\end{align}
The vectors \( \vecbf{z}_1 \) and \( \vecbf{z}_2\) define the position of the dumbbell masses as represented in the asteroid fixed frame and are defined as
\begin{align}
    \vecbf{z}_1 &= R_A^T \parenth{\vecbf{x} + R \vecbf{\rho}_1} , \\
    \vecbf{z}_2 &= R_A^T \parenth{\vecbf{x} + R \vecbf{\rho}_2}, 
\end{align}
where \( \vb{\rho}_i \) defines the position of each mass in the spacecraft fixed body frame.
The gravitational moment on the dumbbell \( \vecbf{M}_i\) is defined as
\begin{align}
    \vecbf{M}_i = m_i \parenth{S(R_A^T \vb{\rho}_i) R^T \deriv{U}{\vb{z}_i}}.
\end{align}
The control inputs to the spacecraft are defined by \( \vb{u}_f, \vb{u}_m \) which define the control force represented in the inertial frame and the control moment represented in the spacecraft frame, respectively.

\subsection{Itokawa Shape Model and Simulated Imagery}\label{sec:imagery}
\begin{figure}[htbp]
    \centering
    \includegraphics[width=0.5\textwidth, keepaspectratio]{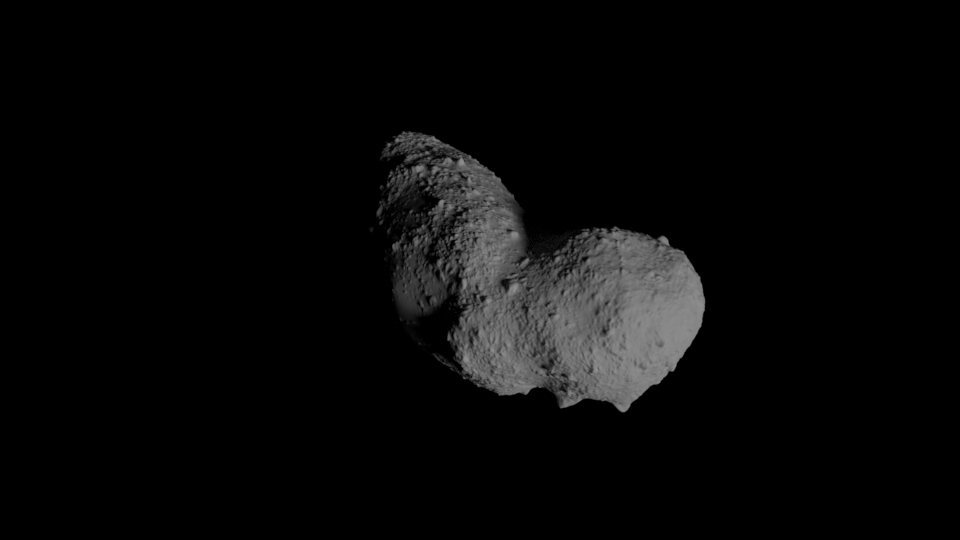}
    \caption{Blender Rendering of Asteroid 25413 Itokawa\label{fig:itokawa_3d}}
\end{figure}

In this work, we consider trajectories about asteroid 25413 Itokawa.
Itokawa was the target of the Hayabusa mission and detailed shape and surface maps have been generated~\cite{kawaguchi2006,tanimoto2013,fujiwara2006}.
We use the estimated rotation period of \SI{12.1}{\hour} with a nominal density of \SI{1.9}{\gram\per\centi\meter\cubed} in the polyhedron potential model.
The shape model is composed of \num{786432} triangular faces and a rendering of the asteroid is provided in~\cref{fig:itokawa_3d}.
A highly detailed model is used for the shape of the asteroid to provide a more detailed and feature rich imaging target. 
However, the polyhedron potential model uses a much coarser shape composed of \num{64} faces. 
This greatly reduces the complexity of the potential model without a significant difference in the qualitative nature of the dynamic environment.

Images of asteroid Itokawa are simulated using Blender, a free and open-source computer graphics software program~\cite{blenderfoundation1995--2017}.
Blender is primarily used by the animation and computer modeling fields to create computer generated images, videos and animations for films and video games. 
Blender offers the capability of accurately modeling the various effects of camera, lighting, and surface properties of a scene using the \texttt{cycles} path-tracing rendering engine.
A unique feature is the ability to compile the Blender rendering software as a Python module.
This allows one to use the capabilities of Blender through a Python API in personal scripts and functions programmatically, rather than through a desktop based graphical interface.
A camera is simulated within Blender with the parameters shown in~\cref{tab:camera_parameters}, which are chosen to emulate the primary sensor of the NEAR spacecraft~\cite{hawkins1997}.
\begin{table}
    \centering
\begin{tabular}{llr}  
\toprule
Parameter & Description &  Value\\
\midrule
FOV & horizontal field of view & \SI{2.25}{\degree}       \\
    &    vertical field of view  & \SI{2.90}{\degree}\\
Image Size & horizontal & \SI{537}{\px}\\
           & vertical & \SI{244}{\px}\\
\(f\)& focal length & \SI{167.35}{\milli\meter}\\
\bottomrule
\end{tabular}
\caption{Camera parameters used in simulation~\label{tab:camera_parameters}}
\end{table}

\section{Nonlinear Landing Controller on \SE}\label{sec:controller}

A wide variety of control schemes have been proposed for asteroid landing missions~\cite{furfaro2013,li2011a}.
In addition, there are a  variety of controllers developed for systems evolving on \( \SE \)~\cite{lee2010,lee2013}.
In this paper, we extend their use from quadrotor aerial vehicles into the space domain. 
This approach addresses many of the issues associated with the related work on asteroid landings.
The geometric control methods used to develop these nonlinear controllers allow for the development of control systems for dynamic systems which evolve on nonlinear manifolds. 
By developing the control system directly on the nonlinear manifold, geometric control techniques provide unique advantages as compared to those developed using local coordinate representations.
Furthermore, the geometric controller avoids the chattering issues inherent in the previous sliding mode control approaches to asteroid landing.
In addition, rather than offering only a bounded stability guarantee, the proposed nonlinear geometric controller guarantees almost global tracking of the attitude and translational states. 
This stability guarantee is critical for mission operations passing close to the surface over highly irregular terrain.
Furthermore, the coupled geometric controller explicitly considers the attitude coupling of the body in contrast to many of the previous approaches.
We briefly summarize the key developments of the \( \SE \) control scheme and leave the detailed derivations to the source manuscripts~\cite{lee2010,lee2013}.

In order to determine the attitude control input, we first define a desired attitude tracking command.
An arbitrary smooth attitude tracking command \( R_d (t) \in \SO \) is given as a function of time.
The corresponding angular velocity command is obtained using the attitude kinematics equation, \( \hat{\Omega}_d = R_d^T \dot{R}_d \).
With the desired attitude command, we then define the errors associated with the attitude and angular velocity.
The attitude and angular velocity tracking errors must be careful chosen to remain on the tangent bundle of \(\SO\).
First, an attitude error function is defined on \( \SO \times \SO \) as
\begin{align}\label{eq:attitude_error_function}
    \Psi(R, R_d) = \frac{1}{2} \tr{I - R_d^T R}.
\end{align}
This positive definite function parameterizes the error between the current attitude, \( R \), and the desired attitude command \( R_d \).
Using the variations of \( \Psi \) gives the attitude tracking error vector \( e_R \in \R^3 \) as
\begin{align}\label{eq:attitude_error_vector}
    e_R = \frac{1}{2} \parenth{R_d^T R - R^T R_d^\vee}.
\end{align}
After further manipulation and using the attitude kinematics equation from~\cref{eq:attitude_kinematics}, it is possible to define the angular velocity tracking error \( e_\Omega \in \R^3 \) as
\begin{align}\label{eq:angular_velocity_error_vector}
    e_\Omega = \Omega - R^T R_d \Omega_d.
\end{align}
With the properly defined attitude error vectors the rotational control input is defined as 
\begin{align*}\label{eq:rotational_control}
    \vb{u}_m = - k_R e_R - k_\Omega e_\Omega + \Omega \times J \Omega - J \parenth{\hat{\Omega} R^T R_d \Omega_d - R^T R_d \dot{\Omega}_d} - \vb{M}_1 - \vb{M_2} 
\end{align*}
where \( k_R, k_\Omega \) are positive controller constants.

The translational control input is defined in a similar manner. 
First we define a smooth tracking command \( x_d(t) \in \R^3 \), which defines the desired position of the spacecraft in the inertial frame.
The tracking error vectors are easier to define as they evolve on a Euclidean space rather than a nonlinear manifold and are given by
\begin{align}
    e_x = x - x_d ,\\
    e_v = v - \dot{x}_d.
\end{align}
With the error variables, the translational control input is then given by
\begin{align}\label{eq:translation_control}
    \vb{u}_f = - k_x e_x  - k_v e_v + ( m_1  + m_2 ) \ddot{x}_d - \vb{F}_1 - \vb{F}_2 ,
\end{align}
where \( k_x, k_v \) are positive constants. 
The control gains are chosen based on the desired closed-loop system response. 
A variety of techniques are available to choose these gains, but a simple linear analysis offers a straightforward and systematic approach to choosing suitable values. 
We use the control inputs defined in~\cref{eq:translation_control,eq:rotational_control} and substitute them into the dynamic equations of motion in~\cref{eq:translational_dynamics,eq:attitude_dynamics}.
This results in the dynamics of the error variables and the gains are chosen to ensure the error behavior meets desired performance criteria, such as percent overshoot or settling time~\cite{nise2004}.

\subsection{Estimating motion from Monocular Imagery}\label{eq:image_processing}

Typically, spacecraft missions require extensive interaction from ground based human operators. 
This interaction ranges from system health checks to navigation and hardware commands.
In addition, there is frequently a large group of analysts in support of any given mission. 
A wide variety of factors make human in the loop control of spacecraft especially difficult.
First, the vast distances cause significant time delays which render it impossible to react immediately to events experienced by the spacecraft. 
Furthermore, deep space missions are designed for continuous mission operations for many years or even decades. 
It is becoming increasingly difficult to maintain trained and knowledgeable staff for several decades in order to support a single mission.
In addition, these operators become increasingly scarce as the contemporary hardware and software tools surpass those of these decades old spacecraft. 
As a result, there is a large focus on completely autonomous spacecraft systems.

\begin{figure}[htbp]
    \centering
    \includegraphics[width=1\textwidth,keepaspectratio]{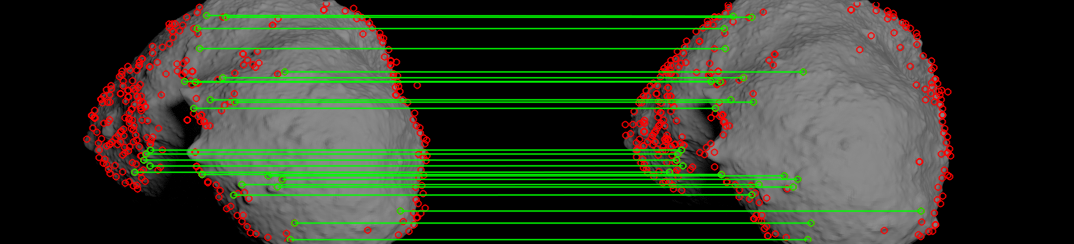}
    \caption{Demonstration of feature detection and matching between simulated images of asteroid 25413 Itokawa\label{fig:itokawa_feature_matching}}
\end{figure}

We seek to utilize well-proven methods in the computer vision and robotics community to first localize the position of the spacecraft from visual imagery.
A secondary step, which is left to a subsequent publication, is then to use this imagery to autonomously update the position of the spacecraft while simultaneously mapping the surface. 
We utilize the state of the art ORB-SLAM implementation to estimate the state of the spacecraft using monocular images~\cite{mur-artal2015}.
This method provides a feature based monocular SLAM system that can operate in real-time for a wide variety of environments. 
ORB-SLAM builds on proven methods in the robotics community to create a custom SLAM system that provides for autonomous tracking and mapping in an unknown environment. 
The first step in ORB-SLAM is to determine accurate features within an image that are most easily recognized in subsequent images. 
This ``feature-extraction'' stage uses the FAST feature detector to determine suitable features, and then the ORB feature descriptor to a store a compact, scale-invariant description of the feature properties.
Next, these features are matched to subsequent images in a ``feature-matching'' stage to determine correspondance between subsequent images. 
For example,~\cref{fig:itokawa_feature_matching} shows a demonstration of features computed in two view of asteroid Itokawa. 
In addition, the correspondance is accurately computed between matching features as shown by the horizontal lines across both images.
From these correspondences, the pose of the camera system is estimated and predicted forward in time using a constant velocity model.
In a simultaneous thread, the ORB-SLAM system builds a local map and stores feature data in order to perform loop closure in the event of returning to a previously imaged location.
We utilize ORB-SLAM to provide an estimate of the position of the rigid spacecraft and will use this estimate to compute the control input.
\section{Numerical Simulation}\label{sec:simulation}

We present a numerical simulation of a rigid dumbbell about asteroid Itokawa.
The dumbbell spacecraft is composed of two equal masses, \( m_1, m_2 = \SI{500}{\kilo\gram} \), separated by \( l = \SI{3}{\meter} \).
The dumbbell body frame is defined with the first body fixed axis, \( \vb{b}_1 \), originating at the center of mass of the spacecraft and directed along the vector from \( m_1 \) towards \( m_2 \).
The other two axes of the spacecraft fixed frame are chosen orthogonal to the \( \vb{b}_1 \) and lie in the plane orthogonal to the dumbbell axis of symmetry. 
A camera, using the parameters from~\cref{tab:camera_parameters}, is aligned with the \( \vb{b}_1 \) axis and used to feed image data to the ORB-SLAM system.
A numerical simulation is used to demonstrate the geometric control of the coupled motion of the spacecraft, and the ability to estimate the motion of the spacecraft from monocular imagery.

The initial condition of the spacecraft is defined as
\begin{align}
    \vb{x}_0 &= \begin{bmatrix} 0 & -2.550 & 0 \end{bmatrix} \si{\kilo\meter}, \\
    R_0 &= \exp {\frac{\pi}{2} \vb{e}_3 }.
\end{align}
The spacecraft begins on the inertial \( \vb{e}_2 \) axis and initially pointing at the asteroid. 
A tracking command is designed to transition the spacecraft towards the asteroid fixed \( \vb{f}_1  \) axis followed by a vertical descent along towards the asteroid surface.
The translational command is divided into two stages, a traverse step where the spacecraft follows a trajectory to align itself with the \( \vb{f}_1 \) axis and a landing step where the spacecraft follows a constant velocity descent towards the surface. 
The desired position command is defined as
\begin{align}
    \vb{x}_d = 
    \begin{cases}
        2.550 \begin{bmatrix} \sin{\omega t} & -\cos{\omega t} & 0 \end{bmatrix}, & t \leq t_d \\
        R_A \begin{bmatrix} \frac{2}{t_d} (t - t_d) + 2.550 & 0 & 0 \end{bmatrix}, & t > t_d , 
    \end{cases}
\end{align}
where \( \omega = \frac{\pi}{2 t_d} \), \( t_d \) is the time from the simulation start when the constant velocity descent should begin, and \( t \) is the simulation time step.

The desired attitude command is chosen such that the spacecraft camera axis, \( \vb{b}_1 \), is directed along the nadir towards the asteroid.
It is sufficient to define two orthogonal vectors to uniquely determine the attitude of the spacecraft.
The \( \vb{b}_{3d} \) vector is chosen to lie in the plane spanned by \(\vb{b}_{1d} \) and \( \vb{e}_3 = \vb{f}_3 \).
The desired attitude command is defined as
\begin{align}
    \vb{b}_{1d} &= - \frac{\vb{x}}{\norm{\vb{x}}} , \\
    \vb{b}_{3d} &= \frac{\vb{f}_3 - \parenth{\vb{f}_3 \cdot \vb{b}_{1d}} \vb{b}_{1d}}{\norm{\vb{f}_3 - \parenth{\vb{f}_3 \cdot \vb{b}_{1d}} \vb{b}_{1d}}}, \\
    \vb{b}_{2d} &= \vb{b}_{3d} \times \vb{b}_{1d} , \\
R_d &= \begin{bmatrix} \vb{b}_{1d} & \vb{b}_{2d} & \vb{b}_{3d} \end{bmatrix}.
\end{align}
The camera axis is aligned with the spacecraft \( \vb{b}_ 1 \) axis, which is directed towards the asteroid, throughout the landing trajectory as the spacecraft moves in the equatorial plane of the asteroid.

\begin{figure}[htbp]
    \captionsetup[subfigure]{position=b}
    \centering
    \subcaptionbox{Planar view of landing trajectory\label{fig:true_landing_trajectory}}{\includegraphics[width=0.5\textwidth]{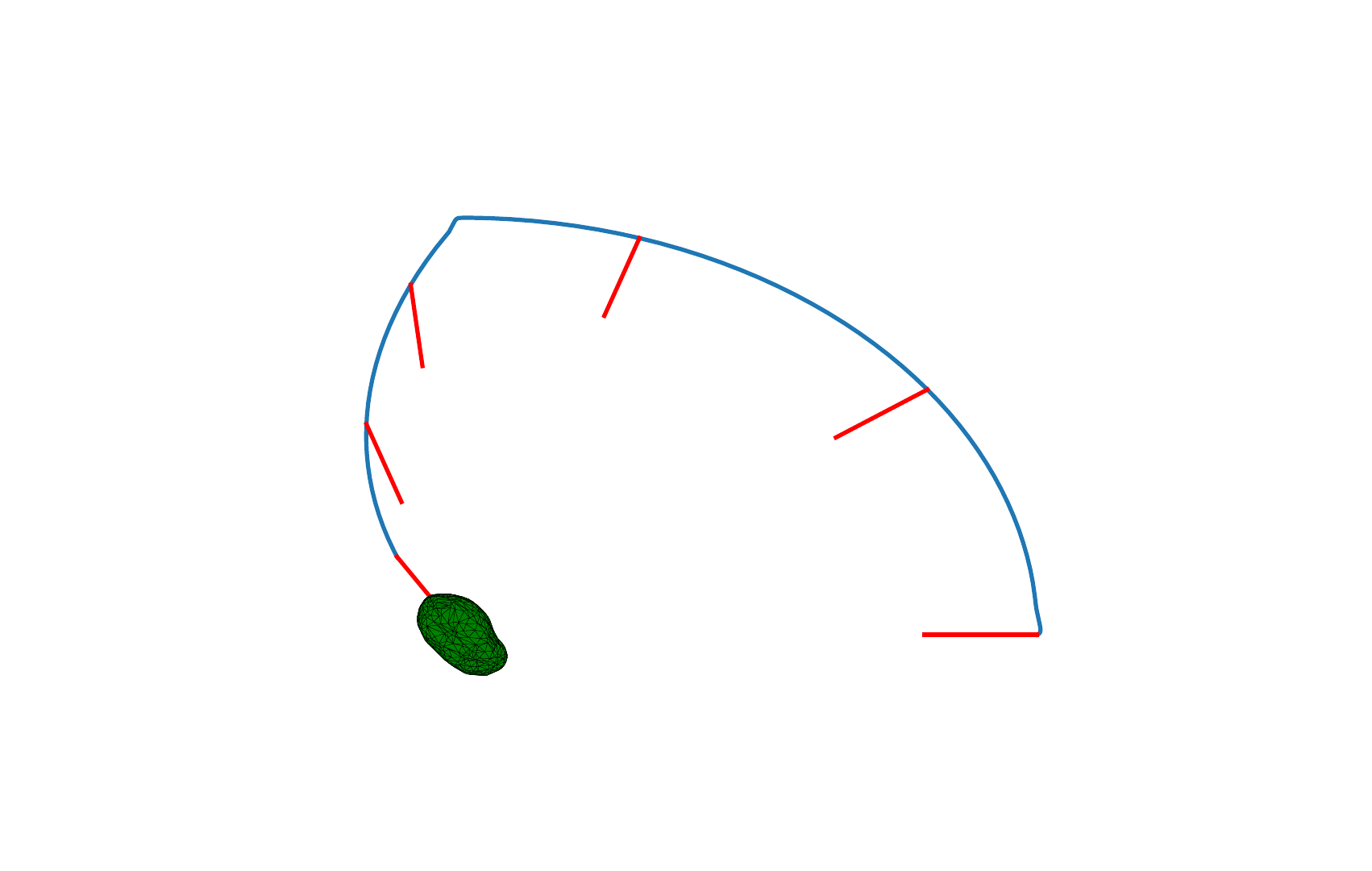}}%
    \subcaptionbox{Position of spacecraft in the inertial frame\label{fig:pos_components}}{\includegraphics[width=0.5\textwidth]{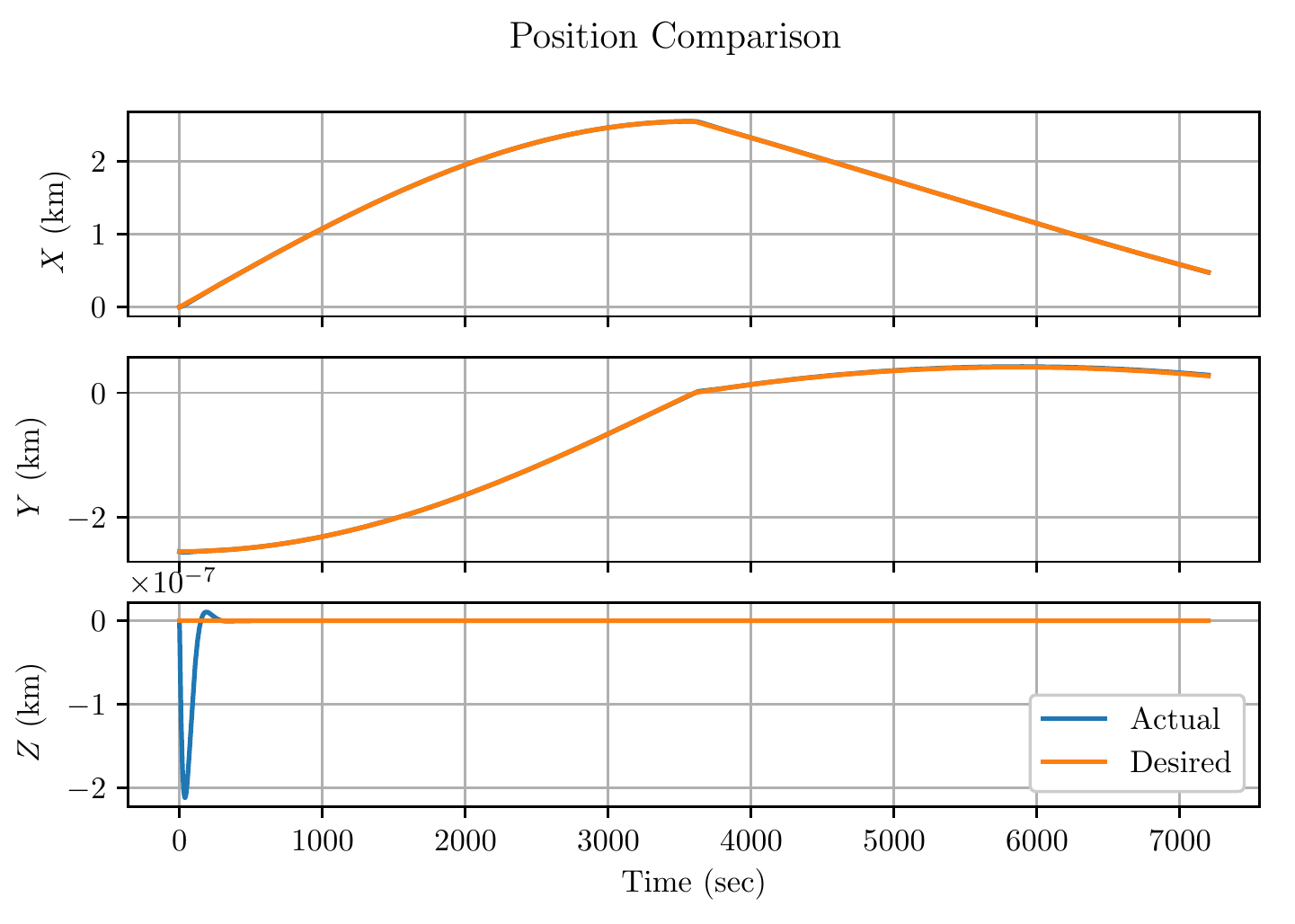}}
    \caption{Landing trajectory to asteroid Itokawa~\label{fig:position}}
\end{figure}

The simulation is carried out over \SI{7200}{\second} with the spacecraft following a circular trajectory for the first \SI{3600}{\sec} before vertically descending in the body fixed frame for the last \SI{3600}{\second}.
\Cref{fig:true_landing_trajectory} shows a view from the positive \( \vb{f}_3 \) pole of the asteroid of the simulated trajectory. 
The position of the center of mass of the dumbbell is shown in blue, while the pointing direction of the camera axis is defined in red. 
The attitude of the dumbbell is displayed at several points along the trajectory demonstrating the pointing of the camera. 
Furthermore, asteroid Itokawa is shown in its final orientation at the completion of the landing simulation. 
\Cref{fig:pos_components} shows that the nonlinear controller is able to accurately track the desired translational trajectory for the duration of the simulation.

\begin{figure}[htbp]
    \centering
    \includegraphics[width=0.75\textwidth,keepaspectratio]{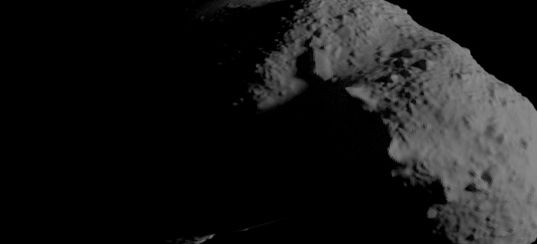}
    \caption{Example image of Itokawa at \( t = \SI{837}{\second}\)\label{fig:example_image}}
\end{figure}

\begin{figure}[htbp]
    \captionsetup[subfigure]{position=b}
    \centering
    \subcaptionbox{Feature tracking visualization\label{fig:orbslam_mappoints}}{\includegraphics[width=0.5\textwidth]{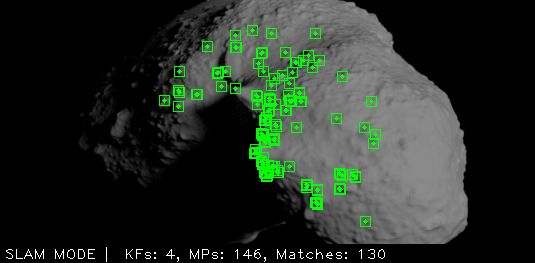}}%
    \subcaptionbox{Localization from monocular images\label{fig:orbslam_localization}}{\includegraphics[width=0.5\textwidth]{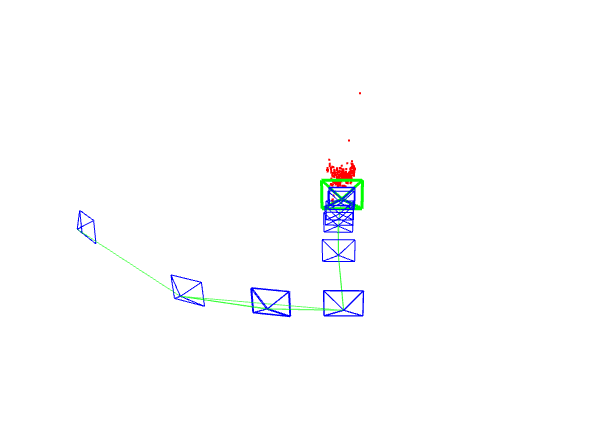}}
    \caption{ORB-SLAM2 feature detection and localization from imagery~\label{fig:orbslam}}
\end{figure}

Concurrent with the simulation, we generate images of Itokawa using Blender, as described earlier.
The images are generated at a rate of \SI{1}{\hertz} and stored for post processing and one example view is shown in~\cref{fig:example_image}. 
From these images we estimate the position of the camera relative to the asteroid using monocular localization.
In this work, we demonstrate the ability to generate realistic imagery and the processing of these images to determine a state estimate of the camera pose.
In future work, we seek to use this state estimate in a closed-loop controller to enable a fully-autonomous control system for the spacecraft. 
From each image, a series of features are detected and tracked between subsequent frames.
A visualization of this process is shown in~\cref{fig:orbslam_mappoints} which shows the tracked features of the image in green.
From these tracked features it is possible to estimate the relative motion of the camera~\cite{szeliski2010}.
\Cref{fig:orbslam_localization} shows a visualization of the localization process from the imagery. 
Through monocular images it is not possible to compute the position of the camera relative to the inertial frame.
Rather, it is only possible to determine a camera pose accurate to an unknown scale factor. 
This limitation is intrinsic to the transformation of the three-dimensional scene onto the two-dimensional image plane~\cite{szeliski2010}.
However, it is possible to compute the relative motion between sequential images and we use this to estimate the motion of the camera.

An additional complexity is due to the fact that the dynamic equations of motion are defined in the inertial reference frame. 
As a result, the asteroid is rotating relative to this frame while the dumbbell is moving relative to the inertial frame. 
The majority of SLAM methodologies use the implicit assumption of a stationary scene or terrain and a moving camera. 
As a result, the motion estimate from the imagery is defined relative to the asteroid fixed frame, rather than the inertial frame. 
However, we assume that the motion of the asteroid is known and constant. 
Over the relatively short timespan of the simulation this is an appropriate assumption and does not significantly affect the solution.
\begin{figure}[htbp]
    \centering
    \includegraphics[width=0.75\textwidth,keepaspectratio]{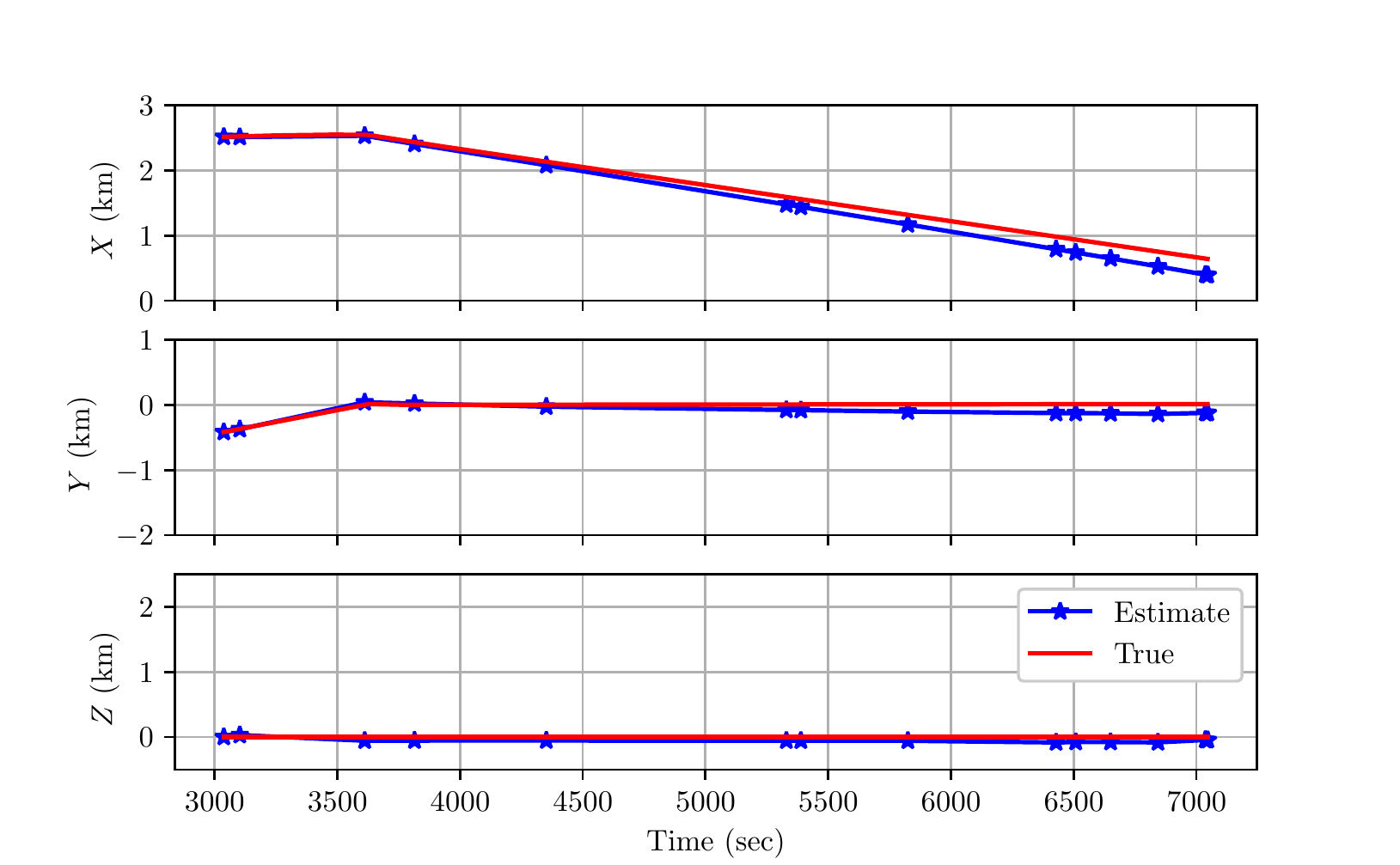}
    \caption{Comparison of image estimate and true trajectory in asteroid fixed frame\label{fig:estimate}}
\end{figure}
\Cref{fig:estimate} shows a comparison of the estimated position in blue and the true position in the asteroid fixed frame. 
The estimate is quite accurate in the \( x \) axis. 
This is primarily due to the fact that during the vertical descent that asteroid remains fixed relative to the spacecraft. 
As a result, a large number of image features are able to be accurately tracked during the motion towards the surface. 
This is also evident by the denser keyframes in the later portion of the simulation as compared to the beginning. 
The estimate tends to diverge in the other two axes.
However, the general trend is captured in the \( y \) axis and the error in the \( z \) axis is on the order of \SI{100}{\meter}. 
This simulation demonstrates the ability to track a desired trajectory for the coupled motion of a rigid spacecraft on \(\SE\).
Furthermore, we show the ability to estimate the motion using a single monocular camera during a short duration landing trajectory. 
In spite of the complex and difficult lighting conditions an estimate is able to be generated.

\section{Conclusions}\label{sec:conclusions}
There have been a variety of approaches for the analysis and design of orbital trajectories around asteroids.
Relatively less work has been directed towards the design of landing trajectories.
Furthermore, much of the previous work has only treated the orbital or translational dynamics.
The approximation of a spacecraft as a point mass rather than an extended rigid body severely limits the applicability and ignores a major component of the dynamic environment.
This work directly derives the equations of motion of a dumbbell spacecraft around an asteroid described using a polyhedron potential model. 
We explicitly consider the impact of the gravitational moment on both the orbit and attitude dynamics. 
With this accurate equations of motion we develop geometric nonlinear controllers which allow for the vehicle to track a desired landing trajectory. 
The desired landing trajectory is accurately followed by the nonlinear controller on \(\SE\).
Furthermore, we use current image processing techniques to estimate the motion of the spacecraft given only two-dimensional images. 
In spite of the challenging environment, with limited lighting and a reduced simulation span, the method is able to accurately track the motion of the vehicle. 
Future work will focus on utilizing this estimate in a closed-loop control and guidance scheme to enable autonomous landing and obstacle avoidance. 
\bibliographystyle{AAS_publication} 
\bibliography{library_local}

\end{document}